\documentclass[usenatbib, usegraphicx]{mn2e} 

\title[Phase models of the Milky Way stellar disc]
{Phase models of the Milky Way stellar disc} 

\author[S.A.~Rodionov, V.V.~Orlov]
{S.A.~Rodionov\thanks{E-mail: seger@astro.spbu.ru} 
and V.V.~Orlov\\
Sobolev Astronomical Institute, 
St. Petersburg State University, 
Universitetskij pr.~28,
198504 St. Petersburg, Stary Peterhof, Russia}           

\date{Accepted ???? ??? ??. Received ???? ??? ??; in original form ???? ??? ??}

\pagerange{\pageref{firstpage}--\pageref{lastpage}} \pubyear{2008}

\begin{document}

\label{firstpage}

\maketitle

\begin{abstract}
We present a new iterative method for constructing equilibrium phase models
of stellar systems. Importantly, this method can provide phase models with
arbitrary mass distributions. The method is based on the following
principle. Our task is to generate an equilibrium $N$-body system with a given
mass distribution. For this purpose, we let the system reach equilibrium
through its dynamical evolution. During this evolution we hold mass
distribution in this system. This principle is realized in our method by
means of an iterative procedure. We have used our method to construct a
phase model of the disc of our Galaxy. In our method, we use the mass
distribution in the Galaxy as input data. Here we used two Galactic density
models (suggested by \citealt{FSLC, DBa}). For a fixed-mass model of the
Galaxy we can construct a 
one-parameter family of equilibrium models of the Galactic disc. We can,
however, choose a unique model using local kinematic parameters that are
known from Hipparcos data. We show that the phase models constructed using
our method are close to equilibrium. The problem of uniqueness for our
models is discussed, and we discuss some further applications of our method.
\end{abstract}


\section{Introduction}

The construction of equilibrium phase models of galaxies is
an important area of research in galactic astronomy. Such models are of
interest from a number of points of view. They are important for
understanding the dynamics of galaxies, and they are necessary for defining
the initial conditions in $N$-body models of stellar systems.

In this paper,
we consider the problem of constructing an equilibrium model of the stellar
disc of a spiral galaxy. Our purpose is to construct a phase model of the
Galactic stellar disc. Various approaches to solving this problem have been
suggested (see, for example, the review in \citealt{RS},
hereafter RS06).

The first approach is based on Jeans equations (equations
for moments of the equilibrium velocity distribution function). One such
method of constructing equilibrium disc models was described by \citet{H}. 
An advantage of this method is its relative simplicity. It is
applicable for a stellar disc with an arbitrary density profile and any
external potential. It has, however, a significant drawback. The system of
Jeans equations used is not closed, so it is necessary to introduce an
additional condition in order to close it. As a result, the constructed
model is often far from equilibrium, as we showed when we used the closure
condition suggested by \citet{H}. A more detailed critical analysis
of this method is given in RS06.

The second approach is based on Jeans
theorem, according to which any function of motion integrals is a solution
of the stationary collisionless Boltzmann equation (see, for example,
\citealt{BT}); that is, it is an equilibrium distribution function. There
is, however, one significant disadvantage of such an approach to
constructing a three-dimensional equilibrium model of a stellar disc. Two
integrals of motion are well known for axisymmetric models: $E$ is energy and
$L_z$ is the angular momentum about the symmetry axis. However, for systems
having phase density $f(E,\,L_z)$, the dispersions of the residual velocities
in the radial and vertical directions have to be the same, which is in
disagreement with observations of spiral galaxies, in particular for the
solar neighbourhood (see, for example, \citealt{DBb}). Axisymmetric
models with different velocity dispersions in the radial and vertical
directions may be constructed if the phase density depends on three
integrals of motion $f(E,\,L_z,\,I_3)$, where $I_3$ is the third integral of
motion. However, an expression for the third integral is not known for the
general case. It is possible to use the energy in vertical oscillations as
the third integral when the residual velocities are much lower than the
rotation velocity with respect to the symmetry axis (cold thin disc). In
such a way, one can construct models of approximately exponential stellar
discs, as done, for example, by \citet{KD, WD}. These authors also describe
the procedure of phase density 
construction for multicomponent models of disc galaxies.

One further
original method for constructing phase galactic models was developed by
\citet{S}. In this method, it is assumed that the total galactic
potential is known. A large number of orbits (library of orbits) in this
potential are constructed, and a model consisting of the particles placed on
these orbits is constructed in such a way that the resulting model has an
initial density profile. We note that this approach is similar to our
Orbit.NB method described below.

In this paper, we use a new iterative
method proposed in RS06. In RS06, the iterative method was applied to
construct a model of the stellar disc of a spiral galaxy (the problem under
consideration). In our notation, this realization of the iterative method is
termed Nbody.SCH. It has, however, a number of disadvantages. The Nbody.SCH
method has a problem with the construction of a relatively cold stellar
disc: using the Nbody.SCH method it is not possible to construct a
sufficiently cold equilibrium stellar disc. The stellar disc of our Galaxy
is rather cold. The model of the stellar disc of our Galaxy constructed
using the Nbody.SCH method is notably far from equilibrium. Moreover, the
Nbody.SCH method cannot be directly applied to a stellar system with
arbitrary geometry (for example, it cannot be applied to elliptical
galaxies).

The first objective of our work is to develop a new realization
of the iterative method without the disadvantages of the Nbody.SCH approach.
The second objective is to construct a phase model of the Galactic disc
using this new method.

Using the iterative method we can construct a phase
model of a stellar system with a given mass distribution. Therefore, in
order to construct a phase model of the Galactic disc we need the mass
distribution of the Galaxy. A number of density models for the Milky Way
Galaxy have been constructed. We use only two of them 
(see \citealt{FSLC, DBa}).

The density models used are described in
Section~\ref{s_densmod}. In Section~\ref{s_itmethod} we present the
iterative method and its
modifications. Two versions of the iterative method are considered in
detail. In our notation, these methods are called Orbit.NB and Nbody.NB. The
Orbit.NB method gives fairly specific and probably non-physical models.
Although such models are probably non-physical, the fact that such
equilibrium models exist is of interest. It gives some insight into the
problem of uniqueness of phase models of the stellar disc. The models
constructed by the Orbit.NB method are discussed in
Section~\ref{s_strange_model}. We think that
the models constructed by Nbody.NB method are physical. In
Section~\ref{s_phase_model} we
present phase models for the Galactic disc constructed using this method. It
is important, that Nbody.NB solves the first objective of our work (see
above). A summary of the results is given in Section~\ref{s_conclusions},
wherein we also 
discuss further applications of our method and models.

\section{Galactic density models}
\label{s_densmod}

There are many Galactic density models in the
literature. We have chosen two of them (see \citealt{FSLC, DBa}).
We note that \citet{DBa} presented a whole family
of density models. We have chosen the model `2' from this paper. Both models
under consideration are axisymmetric.

Let us briefly outline the models used.

\subsection{Model of \citet{FSLC}}
\label{s_densmod_fslc}

This model
contains three main components: a dark halo, central component, and disc.
For the dark halo, the authors used a logarithmic potential 
(see \citealt{BT}, p.\,46):

\begin{equation}
\Phi_H(R,z) = \frac{1}{2} V_H^2 \ln (r^2 + r_H^2) \,,
\end{equation}
where $V_H$ and $r_H$ are the halo parameters
(circular velocity at large $r$ and halo scalelength),
$R$ is the cylindrical radius,
$r = \sqrt{R^2 + z^2}$ is the spherical radius.

The central component
consists of two spherical subsystems. The first one represents the
bulge$+$stellar halo; the second one is the inner
core of the Galaxy. Each component is approximated by a Plummer sphere (see
\citealt{BT}, p.\,42--43). The expression for the whole potential of
the central component has the form
 
\begin{equation}
\Phi_C(R,z) = - \frac{GM_{C_1}}{\sqrt{r^2+{r_{C_1}}^2}} 
- \frac{GM_{C_2}}{\sqrt{r^2+{r_{C_2}}^2}} \,,
\end{equation}
where $G$ is the gravitational constant; 
$M_{C_1}$ and $r_{C_1}$ are the mass and scalelength for the first 
subsystem; 
and $M_{C_2}$ and $r_{C_2}$ are the same parameters for the second.

The disc in this model is the superposition of three
\citet{MN} discs. The whole disc potential has the form

\begin{equation}
\Phi_D (R,z) = - \sum_{n=1}^3 \frac{GM_{D_n}}{ \sqrt{R^2 + ( a_n +
\sqrt{z^2+b^2} )^2 }} \,.
\end{equation}
Here, $b$ is the disc scaleheight (which is the same for all three
components); the parameters $a_n$ are the disc scalelengths; and the values 
$M_{D_n}$ are the masses of the disc components.

\citet{FSLC} have suggested the following values for the above parameters:

\noindent
$r_H = 8.5$~kpc, $V_H = 210$~km\,s$^{-1}$,\\
$r_{C_1} = 2.7$~kpc, $M_{C_1} = 3.0 \cdot 10^9$~$M_\odot$,\\ 
$r_{C_2} = 0.42$~kpc, $M_{C_2} = 1.6 \cdot 10^{10}$~$M_\odot$,\\ 
$b  = 0.3$~kpc,\\
$M_{D_1} = 6.6 \cdot 10^{10}$~$M_\odot$, $a_1 = 5.81$~kpc,\\
$M_{D_2} = -2.9 \cdot 10^{10}$~$M_\odot$, $a_2 = 17.43$~kpc,\\
$M_{D_3} = 3.3 \cdot 10^9$~$M_\odot$, $a_3 = 34.86$~kpc.

\begin{figure*}
\begin{center}
\includegraphics[width=16cm]{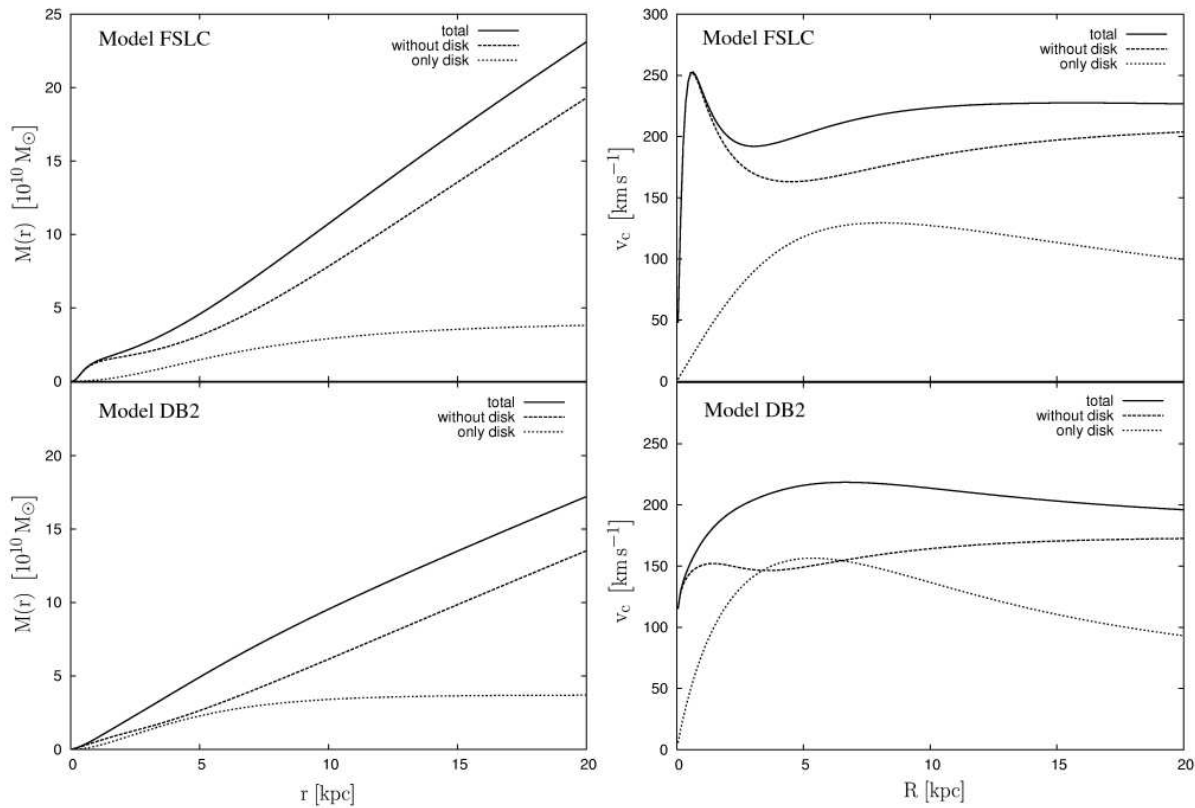}
\end{center}
\caption{
The dependence of cumulative masses on radius (left panels) and
the circular velocity curves (right panels) for various components in the
density models FSLC and DB2. Here the cumulative mass $M(r)$ is the mass
inside the sphere of radius $r$. Solid lines show the dependences for a whole
model; long-dashed lines correspond to the whole model without the disc; and
short-dashed lines correspond to the stellar disc only. We use the thin
stellar disc in the DB2 model as the stellar disc.}
\label{fig_densmod}
\end{figure*}

\noindent
Table~1 in \citet{FSLC} contains a small misprint: instead of
$V_H=220$~km\,s$^{-1}$ it should be $V_H=210$~km\,s$^{-1}$ 
(Flynn, private communication). We note that one of the disc
components ($n=2$) has a negative density; however, the total density in the
disc is positive, and the model is therefore physical. Over a large range in
$R$, the disc density profile is approximately exponential, with a scalelength
of approximately 4~kpc. This is possibly an overestimated value 
(Flynn, private communication). Hereafter, we will refer to this model 
as FSLC. Fig.~\ref{fig_densmod} shows the dependences of cumulative 
masses $M(r)$ and circular velocity
curves for the whole FSLC model, for all components without the disc, and
for the disc only.

\subsection{Model of \citet{DBa}}
\label{s_densmod_db}

In addition to the
FSLC model, we consider one model of the family suggested by \citet{DBa}.
Every model of this family consists of five components. There are
three disc components (interstellar medium (ISM), thin stellar disc, and
thick stellar disc) and two spheroidal components (dark halo and bulge).

The distribution of volume density in each disc component has the form

\begin{equation}
\rho_d(R,z) = \frac{\Sigma_d}{2z_d} \exp 
\biggl(  - \frac{R_m}{R_d} - \frac{R}{R_d} - \frac{|z|}{z_d} \biggr).
\end{equation}
Here $\Sigma_d$ is the central surface density of the component, and
parameters $R_d$ and $z_d$ give scalelength and scaleheight of the
component.
By using the parameter $R_m$ one can introduce a central density depression
in the disk.

The density distribution for each spheroidal component has the form

\begin{equation}
\rho_s(m) = \rho_0 \biggl( \frac{m}{r_0} \biggr)^{- \gamma}
\biggl( 1 + \frac{m}{r_0} \biggr)^{\gamma - \beta} 
\exp{ (- m^2 / r_t^2) },
\end{equation}
where 
\begin{equation}
m = \sqrt{ R^2 + q^{-2} z^2}.
\end{equation}
Here $\rho_0$, $r_0$, $\gamma$, $\beta$, $q$, $r_t$ are the parameters
of the spheroidal components.

We use model 2 from this paper, hereafter the DB2 model. This choice is
somewhat arbitrary. We do not consider other models from this paper because
a comparison of the different Galactic models is outwith the goals of this
paper. The parameters of the DB2 model are shown in
Tables~\ref{tab_db2_disks}~and~\ref{tab_db2_sphs}. The
details of the construction of this model are given in \citet{DBa}.

\begin{table}
\caption{Parameters for three disk components in the DB2~model.}
\label{tab_db2_disks}
\begin{center}
\begin{tabular}{c|cccc}
\hline
comp. & $\Sigma_d, \; {\rm M_{\odot} \, pc^{-2}}$ &
$R_d, \; {\rm kpc}$ & $R_m, \; {\rm kpc}$ & $z_d, \; {\rm kpc}$\\
\hline
thin disk  & 1022  & 2.4 & 0 & 0.18\\ 
thick disk & 73.03 & 2.4 & 0 & 1   \\
ISM        & 113.6 & 4.8 & 4 & 0.04\\
\hline
\end{tabular}
\end{center}
\end{table}

\begin{table}
\caption{Parameters for the two spheroidal components in the DB2~model.}
\label{tab_db2_sphs}
\begin{center}
\begin{tabular}{c|cccccc}
\hline
comp. & $\rho_0, \; {\rm M_{\odot} \, pc^{-3}}$ &
$r_0, \; {\rm kpc}$ & $\gamma$ & $\beta$ & $q$ & $r_t, \; {\rm kpc}$\\
\hline
bulge & 0.7561 & 1     &  1.8 & 1.8   & 0.6 & 1.9 \\ 
halo  & 1.263  & 1.090 & -2   & 2.207 & 0.8 & $\infty$ \\
\hline
\end{tabular}
\end{center}
\end{table}

Here we construct an equilibrium $N$-body model of the stellar
Galactic disc. As the stellar disc, we take only the thin stellar disc from
the DB2 model. The construction of a two-component stellar disc in this
model will be the subject of future investigations.

In addition to the
density, we need to calculate the potentials of the various components in
the DB2 model. The potentials were numerically calculated using the code
GALPOT by Walter Dehnen. The method of potential determination is described
in \citep{DBa}. The code was taken from the NEMO~package
(http://astro.udm.edu/nemo; \citealt{T}).

The cumulative mass profile
$M(r)$ and circular velocity curve for the DB2 model are shown in
Fig.~\ref{fig_densmod} for
the whole model, for all components apart from the thin stellar disc, and
for the thin stellar disc only.

Fig.~\ref{fig_densmod} shows that the FSLC and DB2 models
are different. The FSLC model has a more massive and concentrated bulge with
respect to the DB2 model. In particular, this massive bulge is the reason
for the central peak in the rotation curve for the FSLC model. Furthermore,
the relative disc contribution in the whole mass and the circular velocity
curve for the inner model ($R \le 8 \, {\rm kpc}$) is significantly higher
in the DB2 model  
than in the FSLC model.

\section{Iterative method for equilibrium models constructing}
\label{s_itmethod}

\subsection{Basic idea of iterative method}
\label{s_itbasic}

The iterative method
is used to construct $N$-body models close to equilibrium and with a given
mass distribution (see RS06 for details). The basic idea of this approach is
as follows. First, we generate an $N$-body system with a given mass
distribution but with arbitrary initial particle velocities (which, for
example, can be taken as zero). Furthermore, we let the system reach
equilibrium through its dynamical evolution. During this evolution we hold
mass distribution in the system. If necessary, some parameters of the
velocity distribution can be fixed.
This is achieved in the following way.

The general algorithm of the iterative method is as follows.
\begin{enumerate}
\renewcommand{\theenumi}{(\arabic{enumi})}
\item
An N-body
system with a given mass distribution but with arbitrary particle velocities
is constructed. The velocities can, for example, be taken to be zero.
\item
The system is evolved on a short time-scale.
\item
We construct a new N-body
system, with the same given mass distribution but with velocities chosen
according to the velocity distribution in the system already evolved. We
note that, if there are some limitations on the velocity distribution, this
distribution should be corrected taking into account these restrictions (see
the discussion below).
\item
We return to point 2. We repeat such cycles until
the velocity distribution stops changing.
\end{enumerate}

As a result, we obtain an $N$-body
model close to equilibrium that has a given density profile (see RS06 and
our results below for details).

We can discuss the iterative approach in a
more general manner. When one needs to find an equilibrium state of an
arbitrary dynamical system, but so that this state has some necessary
properties (in the case under study, the dynamical system is a set of
gravitating points and the necessary property is the density profile), one
can simply give a possibility for the system itself to tune to the
equilibrium state, holding the necessary parameters.

The idea of our
iterative method is simple. Its realization in practice is more complicated.
The main difficulty is the third stage, when it is necessary to construct a
model with the same velocity distribution as the evolved model from the
previous iteration step.

Below we discuss an application of the iterative
procedure to the problem of constructing an equilibrium model of the
Galactic disc.

\subsection{Equilibrium models of stellar disks}
\label{s_itdisk}

\subsubsection{Family of equilibrium models}
\label{s_fixLz}

Our task is to construct an equilibrium model of the
stellar disc of our Galaxy. We consider all Galactic components as
axisymmetric, and can formulate our task in the following way. We need to
construct an equilibrium $N$-body model of a stellar disc with a fixed density
distribution $\rho_{\rm disk}(R,z)$ that is embedded in the rigid external
potential  $\Phi_{\rm ext}(R,z)$, where $\Phi_{\rm ext}(R,z)$ is created by all
Galactic components except 
the stellar disc (i.e. the dark halo and bulge).

It can be expected that at
least a one-parameter family of equilibrium models will exist when the
functions $\rho_{\rm disk}(R,z)$ and $\Phi_{\rm ext}(R,z)$ are fixed. The
parameter of this family is the fraction of kinetic energy contained in the
residual motions.

The reason why this family exists is as follows. It is possible to show
that, if $\rho_{\rm disk}(R,z)$ and $\Phi_{\rm ext}(R,z)$ are fixed, then
for all equilibrium 
discs the total kinetic energy should be the same. This is a direct
consequence of the virial theorem. This kinetic energy can, however, be
distributed between regular rotation and residual velocities in different
manners. Cold equilibrium models exist when a large fraction of the kinetic
energy is concentrated in the regular rotation (an extreme case is the model
with circular orbits), and hot equilibrium models exist when a large
fraction of the kinetic energy is concentrated in the residual motions.

In
RS06, the authors showed that, for exponential disc, a one-parameter family
of models can be constructed by the iterative method for fixed $\rho_{\rm
disk}(R,z)$ and $\Phi_{\rm ext}(R,z)$.
If the iterations are started from different initial states,
different models result; however, they all form a one-parameter family. As
expected, the parameter is the fraction of kinetic energy concentrated in
the residual motions. In order to obtain a definite model from this
one-parameter family, the method suggested in RS06 can be used. We simply
fix the fraction of kinetic energy during the iterative process. In
principle, we could fix any value characterizing the ``heat'' degree of the
disc. The authors of RS06 suggest using the value of angular momentum about
the $z$-axis (symmetry axis) as this parameter:

\begin{equation}
L_z = \sum_{i=1}^N m_i v_{\varphi i} R_i \, ,
\end{equation}
where $m_i$, $v_{\varphi i}$, $R_i$ are
the mass, azimuthal velocity, and cylindrical radius
of the $i$-th particle.

We fix the value of $L_z$ at each iteration step.
When we have constructed a new model (which has the same velocity
distribution as the slightly evolved model from the previous iteration
step), we correct the azimuthal velocities so that the total angular
momentum of the system is the same as the fixed value of $L_z$ . This is done
as follows. Let $L_z$ be this fixed angular momentum, and let $L_z^\prime$
be the 
current value of the angular momentum. The new azimuthal velocities of
particles are prescribed as follows:
\begin{equation}
v_{\varphi i} = v_{\varphi i}^\prime + 
\frac{(L_z - L_z^\prime)}{{\sum_{j=1}^{N}R_j m_j}}
 \, ,
\end{equation}
where $v_{\varphi i}^\prime$ is the current value of the azimuthal velocity
of the $i$-th particle, 
and $v_{\varphi i}$ is the corrected azimuthal velocity of the $i$-th
particle.  

We note
that by using the iterative method with fixed $L_z$, one can construct colder
models with respect to the ones without $L_z$ fixed (because the cold stellar
disc tends to heating during the dynamical evolution). The stellar disc of
the Galaxy is rather cold. Thus it is difficult to construct a cold model of
the Galactic disc without fixing $L_z$. We therefore fix $L_z$ in all our
models. 

\subsubsection{Variants of the velocity distribution ``transfer''}

We first discuss the
core of the iterative method, namely an algorithm to transfer the velocity
distribution (item 3 in the iterative procedure).

The transfer problem is as
follows. We have an ``old'' model. This is an evolved model from the previous
iteration step from which we would like to copy a velocity distribution.
Moreover, we have a ``new'' model that is constructed according to the fixed
density distribution. We have to give the velocities to the particles in the
new model using the velocity distribution in the old model. How do we do
this?

In RS06, the authors used an algorithm of velocity distribution
``transfer'', which is based on assuming that the particles have a truncated
Schwarzschild velocity distribution. We describe this approach briefly. We
take a disc model (old model) from which we
are going to ``copy'' the velocity distribution. The model is divided along
the $R$-axis into various regions (concentric cylindrical tubes). For each
region, we calculate four velocity distribution moments ($\bar v_{\varphi}$,
$\sigma_R$, $\sigma_{\varphi}$, $\sigma_z$~--- the
mean azimuthal velocity and three dispersions of residual velocities along
the directions $R$, $\varphi$, and $z$). These moments are used for the
velocity choice 
in the new model. We assume that the velocity distribution is the
Schwarzschild one, but without the particles that can go out of the disc
(see RS06 for details).

We slightly modified this scheme of velocity
transfer. The model is divided into regions not only along the $R$-axis, but
also along the $z$-axis. The regions are chosen in such a way that they all
contain similar numbers of particles.

This method of velocity distribution
transfer has, however, two drawbacks (even in its modified form). First, an
{\it a priori} assumption is made that the velocity distribution is the
Schwarzschild one. Second (and more importantly), this method cannot be used
for other geometry systems (e.g. triaxial elliptical galaxies).

We have
developed another method of velocity distribution transfer. We believe that
it is a more general and simpler method. The basic idea of this new method
is as follows. We prescribe to the new-model particles the velocities of
those particles from the old model that are nearest to the ones in the new
model.

The simplest (although not quite successful, as we show below)
implementation of this idea is evident. One can prescribe to each particle
in the new model the velocity of the nearest particle from the old model.
Let us formulate this proposition more strictly. For each $i$th particle from
the new model, one finds the old-model $j$th particle with the minimum value
of $|{\bf r}^{new}_i - {\bf r}^{old}_j|$. 
Here, ${\bf r}^{new}_i$ is the radius vector of the $i$th particle
in the new model, and ${\bf r}^{old}_j$ is the radius vector of the $j$th
particle in the 
old model. One then takes as the velocity of the $i$th particle in the new
model the velocity of the $j$th particle in the old model.

This simple
algorithm has, however, one essential problem. If the numbers of particles
in the old and new models are the same then only about one-half of the
particles in the old model participate in the velocity transfer. The reason
is that many old-model particles have a few particles in the new model to
which they transfer their velocities. At the same time, almost one-half of
the particles in the old model do not transfer their velocities at all. This
means that a significant amount of information on the velocity distribution
will be lost in the transfer process. The noise will therefore grow, and
this is indeed observed in numerical simulations. Thus, one cannot construct
an $N$-body model close to equilibrium by the iterative method described above
using this transfer algorithm.

It is, however, possible to modify the
transfer scheme in order to overcome this failure. An input parameter of
this improved algorithm is the ``number of neighbours'' $n_{nb}$ . For each
particle in the old model, we introduce the parameter $n_{use}$, which denotes
the number of times this particle is used for velocity copying. At the
beginning of the transfer procedure, we assume $n_{use}=0$ for each particle in
the old model. We then consider all particles in the new model. For each
particle from the new model, we find the nearest $n_{nb}$ neighbours in the old
model (in this case, the closeness is understood as the minimum distance
between the particles in the old model and the point at which the particle
of interest from the new model is placed). We then reveal a subgroup of
particles that have a minimum $n_{use}$ among these $n_{nb}$ neighbours, and among
this subgroup we find the particle that is the closest one to the position
of the new-model particle. We prescribe to the new-model particle the
velocity of the found particle from the old model. Moreover, we add one to
the parameter $n_{use}$ of this old-model particle.

We note that this algorithm is the same as the previous one if $n_{nb}=1$. As
mentioned above, the problem with this algorithm is that only one-half of
the particles take part in the velocity distribution transfer. If we take
$n_{nb}=10$, only a small fraction (a few per cent) of old-model particles do
not take part in the transfer process. Using this improved transfer method
in the iterative procedure gives good results in the sense that the
constructed models are close to equilibrium.

In this method, one can take
into account that the galactic models under study are axisymmetric. One just
has to redefine the distance between the particles. That is, one can search
for the nearest particles in two-dimensional space $Rz$ instead of
three-dimensional space $XYZ$. In this case, one should transfer the
velocities in cylindrical coordinates in order to remove any dependence on
the azimuthal angle. This guarantees that the constructed model has an
axisymmetric velocity distribution. We use this method when constructing the
model of the Galactic stellar disc (see below).

We note that this method of
velocity transfer is universal, and can be applied in systems with arbitrary
geometry. The models constructed in this way are close to equilibrium, but,
partly because of this, the method has a small disadvantage. The iterations
converge much more slowly than the ones in the above-mentioned method based
on the Schwarzschild velocity distribution. The reason for this slow
convergence is that even intermediate models are fairly close to
equilibrium, so that the models change only slightly over one iteration, and
the iterations converge slowly.

\subsubsection{Different ways of system evolution simulations}
\label{s_evolv_method}

We
discuss one more way to modify the iterative method. In the general
algorithm of the iterative procedure, item~2 allows the model to evolve over
a short time-scale. This means that one needs to simulate the
self-consistent $N$-body evolution of the stellar disc during a short time in
the field of the external potential $\Phi_{\rm ext}$. 
There is, however, another
possibility. Instead of simulating the self-consistent dynamical evolution
of the $N$-body system, one can simulate the motions of $N$ massless test
particles in the regular galactic potential $\Phi_{\rm disk} + \Phi_{\rm ext}$,
where $\Phi_{\rm disk}$ is the
disc potential corresponding to the density $\rho_{\rm disk}$.
We note that the
simulation of a system of $N$ test particles in a rigid potential is much less
cumbersome than simulation of the self-consistent $N$-body system.

At first
glance, both methods have to give practically the same results, because the
initial stellar disc has the density profile $\rho_{\rm disk}$ that creates the
potential $\Phi_{\rm disk}$. It is expected that the iterations will
converge to an 
equilibrium state. Therefore in the self-consistent case, the disc at the
late stages of the iterative process will be close to equilibrium and will
not significantly change its density profile during a single iteration
(especially because we follow the evolution over a short time-scale).

However, the iterative methods using these two modes of calculation may lead
to essentially different results (see below).

\subsubsection{Comparison of different realizations of iterative method}

We thus have two ways to perform the
velocity distribution transfer. The first one is based on calculations of
the velocity distribution moments and assumes a Schwarzschild velocity
distribution (hereafter we refer to this method as SCH). The second is based
on prescribing to the particles in the new model the velocities of the
nearest particles in the old model (hereafter we refer to this method as
NB).

We also have two ways of performing system simulations. The first one
is the self-consistent simulation of the N-body gravitating
system evolution (hereafter we refer to this approach as Nbody). The second
is the calculation of massless particle orbits in the regular potential
$\Phi_{\rm disk} + \Phi_{\rm ext}$ (hereafter we refer to this approach as
Orbit). 

We thus have four
versions of the iterative method: Nbody.SCH, Nbody.NB, Orbit.SCH, and
Orbit.NB. Our task is to choose from among them the method that will be used
to construct a phase model of the Galactic stellar disc.

We will show in
Section~\ref{s_strange_model} that the Orbit.NB approach gives models that
are probably 
non-physical. However, the very fact of the existence of such ``strange''
equilibrium models is of interest, and gives food for thought (see
Section~\ref{s_strange_model}
for details). 

We thus need to choose from among three approaches: Nbody.NB,
Nbody.SCH, and Orbit.SCH. Our test simulations have shown the following. The
models of hot discs constructed by all these approaches are similar. The
only exception is that the models constructed by the Nbody.NB method are
slightly closer to equilibrium. However, the models of cold discs are
significantly different. In particular, this concerns models of the Galactic
disc, because the Galactic disc is rather cold. The models of cold discs
constructed with the Nbody.NB approach are close to equilibrium, whereas the
ones constructed with the Orbit.SCH and Nbody.SCH approaches are quite far
from equilibrium. We can therefore conclude that the Orbit.SCH and Nbody.SCH
approaches are not suitable for constructing equilibrium models of cold
stellar discs. Moreover, from a methodical point of view, it is more correct
to use the NB transfer approach, because here we do not make any 
{\it a priori}
assumptions concerning the form of the velocity distribution (in contrast to
the SCH method).

We shall thus use the Nbody.NB approach to construct phase
models of the Galactic stellar disc (see Section~\ref{s_phase_model}).

\subsubsection{Technical comments}
\label{s_tmoments}

We note a few important technical details. 

In the iterative method,
there is one parameter~--- the time interval $t_i$ of each iteration. This is the
time interval over which the system evolves during each iteration. How do
we choose the value of $t_i$? This time should not be too short, because in
this case the system would have no time to evolve during one iteration step.
On the other hand, it should not be too long. At least, this time should be
much shorter than the typical times of development of various instabilities;
otherwise, these instabilities may change the system substantially. We
cannot suggest any strict criterion for choosing $t_i$. We have chosen a value
from numerical simulations. Our simulations have shown that, if we take $t_i$
within reasonable limits (not too short and not too long), the constructed
models are the same (within the noise limits). Of course, this is valid only
when we use the model with the same $L_z$ (see Section~\ref{s_fixLz}). In
any case, a 
basic test of every method to construct the equilibrium models would be a
numerical check that the model is close to equilibrium.

We have used the
following modification of the iterative method in several simulations. We do
not take a fixed iteration time, but choose this time randomly within the
range $(0,\,t^{\rm max}_i)$. If one makes the iterations with the 
fixed $t_i$, the
following situation may occur in principle. The iterations could converge to
an artificial non-equilibrium state when the model has large changes at
intermediate times within one iteration, however in the end of iteration, it
could have the same state as in the beginning of the iteration. Another
situation that could occur is that the model could jump from one state to
another, i.e. oscillations between two states occur. The random choice of
the iteration step enables us to avoid such situations. However, if we
consider the Nbody.NB approach, fixed and random iteration steps give
practically the same models as output.

In many of our simulations, we used
the following method to decrease the CPU time. Initially, we make the
iterations have a low accuracy (for example by using a smaller number $N$ of
bodies) and then gradually increase the accuracy up to a pre-defined limit.
This scheme was used in all simulations of Section~\ref{s_phase_model}.

In all our $N$-body
simulations (self-consistent scheme), we used the TREE code \citep{BH} 
and a few other codes from the NEMO package
(http://www.astro.umd.edu/nemo; \citealt{T}). We used our original codes to
simulate the motions of particles in the rigid potential.

\section{Non-physical models. hypothesis of uniqueness}
\label{s_strange_model}

\subsection{Models constructed with the Orbit.NB approach}
\label{s_strange_model_s1}

In this section we consider the
models constructed with the Orbit.NB approach. We will show that these
models are probably non-physical models. However, the very fact of the
existence of such models is of interest. It gives some insight into the
problem of uniqueness of phase models of the stellar disc.

One noteworthy
feature of the iterative Orbit.NB approach is that usually at fixed
$\rho_{\rm disk}(R,z)$, $\Phi_{\rm ext}(R,z)$, and $L_z$ essentially
different models can be constructed by
means of this approach, although all other versions of the iterative method
give similar (within the limits of noise) final models at a fixed value of
$L_z$. Another important feature of this approach is that the velocity
distributions of the final models are very different from the Schwarzschild
distribution. We consider this fact in more detail below.

Let us consider a
model constructed by the Orbit.NB approach. We take the FSLC density model
(see Section~\ref{s_densmod_fslc}); that is, the disc density $\rho_{\rm
disk}$ is taken from the FSLC 
model, and the rigid potential $\Phi_{\rm ext}$ is generated by all FSLC
model components 
except the disc. The disc density is taken to be zero at 
$R > R_{\rm max} = 30 \; \rm kpc$ or $|z| > z_{\rm max} = 10 \; \rm kpc$. 
We take the cold initial model in which all
particles move along circular orbits. We used 1000 iterations for 
$N = 200,000$, and then 100 iterations for $N = 500,000$. 
The integration step was taken
as $dt = 0.5 \; \rm Myr$. The number of neighbours for the velocity 
distribution
transfer was taken to be $n_{\rm nb} = 100$. We used a scheme with randomized
iteration time (see Section~\ref{s_tmoments}) with 
$t_i^{\rm max} = 100 \; \rm Myr$. During the
iterative process, we fixed the angular momentum as 
$L_z = 6.412 \cdot 10^{13} \; M_{\odot} \: \rm kpc  \: km \: s^{-1}$.
We note that in all simulations the following system of units was
used: gravity constant $G=1$, length unit $u_r = 1 \; \rm kpc$,
time unit $u_t = 1 \; \rm Myr$. 
In this system of units, the chosen value is  $L_z=0.295$ (hereafter we
indicate the parameter $L_z$ in this system of units). We refer to this model
as FSLC.O. We consider only this model; however, we emphasize that very
different models may be constructed by the Orbit.NB approach at fixed $L_z$ if
we take different initial models.

In addition to the constructed FSLC.O
model, we consider its changes during further dynamical evolution
(self-consistent $N$-body evolution of the constructed stellar disc 
in the field of the external potential $\Phi_{\rm ext}$). The
parameters of simulation are taken as follows: number of bodies $N=100\,000$,
integration time step $dt=0.04 \; \rm Myr$, softening length 
$\epsilon=0.025 \; \rm kpc$. Two last
parameters were chosen according to the recommendations of \citet{RS05}.

\begin{figure}
\begin{center}
\includegraphics[width=80 mm]{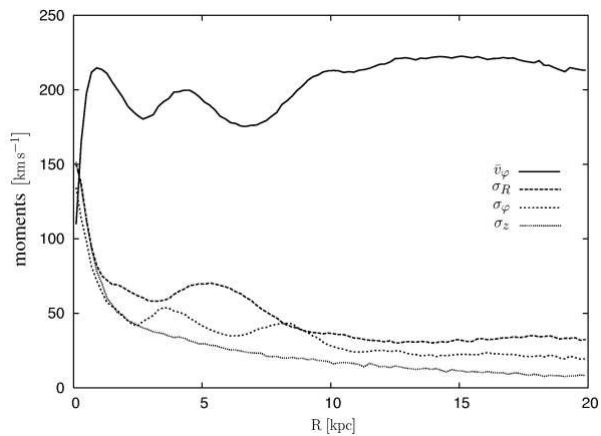}
\end{center}
\caption{
Dependences of $\bar v_{\varphi}$, $\sigma_R$,
$\sigma_{\varphi}$, and $\sigma_z$ on $R$ in the FSLC.O model
constructed by the Orbit.NB~method.
}
\label{fig_FSLC.O_moments}
\end{figure}

\begin{figure*}
\begin{center}
\includegraphics[width=16cm]{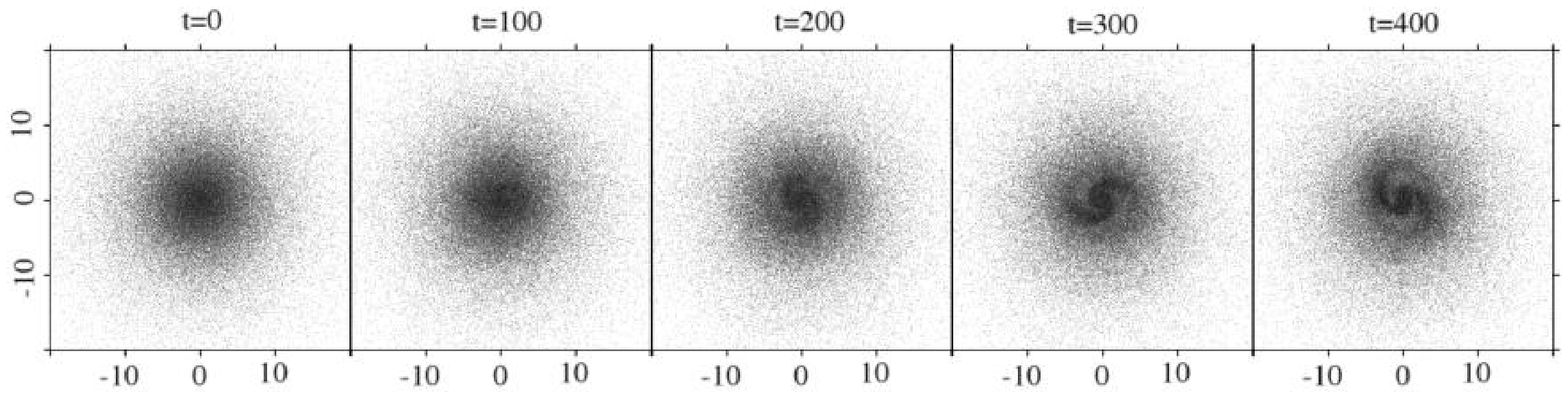}
\includegraphics[width=16cm]{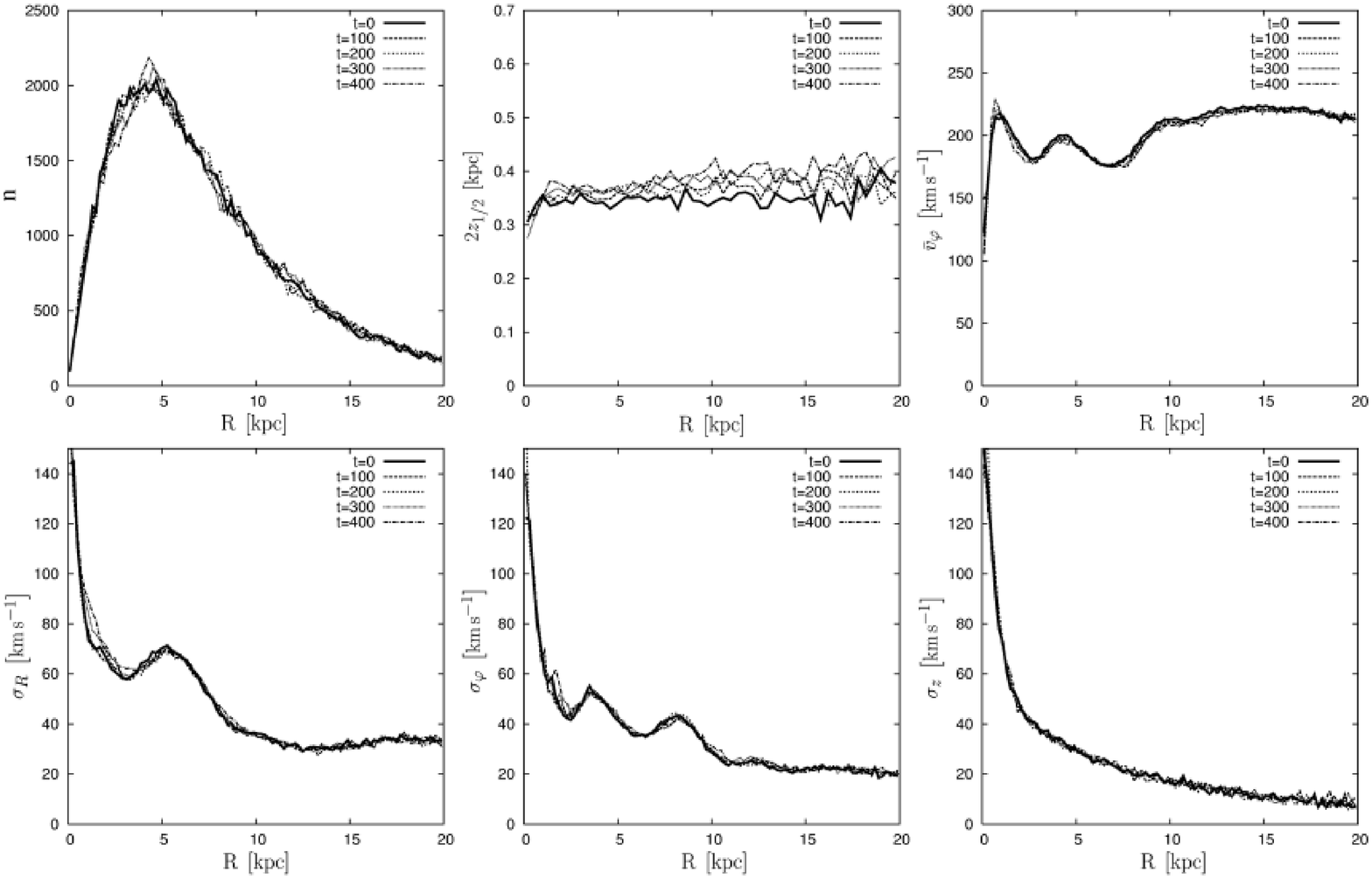}
\end{center}
\caption{
Initial evolutionary stages for the FSLC.O model. The upper
snapshots show the disc views face-on for various moments of time (0, 100,
200, 300 and 400~Myr); the grey intensities correspond to the logarithms of
particle numbers in the pixels. Middle and bottom panels show the
dependences of various disc parameters on the cylindrical radius $R$ for
various moments of time. Here $n$ is the number of particles in concentric
cylindrical layers; $2 z_{1/2}$ is twice the median of the value $|z|$ (it is a
parameter of the disc thickness, see \citealt{SR}); $\bar v_{\varphi}$,
$\sigma_R$, $\sigma_{\varphi}$, $\sigma_z$ are four moments of the velocity
distribution.}
\label{fig_FSLC.O_stability}
\end{figure*}

The radial profiles for four moments of the velocity
distribution $\bar v_{\varphi}$, $\sigma_R$,
$\sigma_{\varphi}$, and $\sigma_z$, are shown in
Fig.~\ref{fig_FSLC.O_moments}. It can be seen that
the profiles of $\bar v_{\varphi}$, $\sigma_R$,
and $\sigma_{\varphi}$ have unusual forms. They have various peaks
and dips. It would seem that such a complicated system cannot be stable;
however, the FSLC.O model is close to equilibrium! When we follow its
evolution, it emerges that the constructed model conserves structural and
dynamical parameters very well (see Fig.~\ref{fig_FSLC.O_stability}).

\begin{figure*}
\begin{center}
\includegraphics[width=16cm]{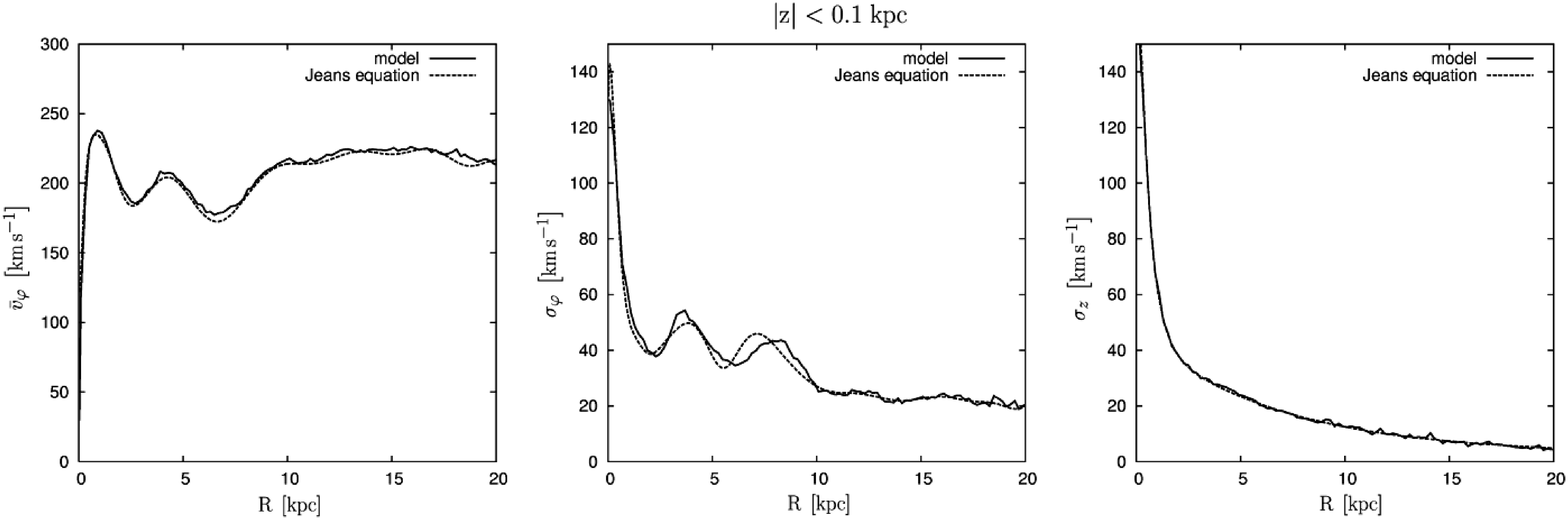}
\end{center}
\caption{
Comparison of profiles of the velocity distribution moments
calculated from the Jeans equations and from the FSLC.O model. All moments
for the disc were calculated inside the region $|z| < 0.1 \; {\rm kpc}$.
Left panel: 
the solid line corresponds to the value $\bar v_{\varphi}$ for the model,
and the dashed line 
corresponds to the same value calculated from the Jeans equation (the
first equation of the system (\ref{eq_jeans})), where the values $\sigma_R$
and $\sigma_{\varphi}$
were taken from 
the model. Middle panel: the solid line corresponds to the value
$\sigma_{\varphi}$ for the
model, and the dashed line corresponds to the same value calculated from the
Jeans equation (the second equation of the system (\ref{eq_jeans})),
where the values $\bar v_{\varphi}$ and $\sigma_R$ were taken from the model. 
Right panel: the solid line corresponds to
the value $\sigma_z$ for the model, 
and the dashed line corresponds to the same
value calculated from the Jeans equation (the third equation of the system
(\ref{eq_jeans})).}
\label{fig_FSLC.O_jeans}
\end{figure*}

An interesting question arises
in connection with the FSLC.O model equilibrium: how do the moments of the
velocity distribution satisfy the equilibrium Jeans equations (see
\citealt{BT}):

\begin{equation}
\label{eq_jeans}
\left\{
\begin{array}{rcl}
{\overline v}_{\varphi}^2 &=& v_{\rm c}^2 + \sigma_R^2  -
\sigma_{\varphi}^2 + \displaystyle \frac{R}
{\rho_{\rm disk}}
\frac{\partial (\rho_{\rm disk} \sigma_R^2)}{\partial R} \, , \\
\sigma_{\varphi}^2 &=& \displaystyle\frac{\sigma_R^2 R}
{2 \overline v_{\varphi}}
\left( \frac{\partial \overline v_{\varphi}}{\partial R} +
\frac{\overline v_{\varphi}}{R} \right) \, , \\
\displaystyle\frac{\partial (\rho_{\rm disk} \sigma_z^2)}{\partial z} &=&
-\rho_{\rm disk}
\displaystyle\frac{\partial \Phi_{\rm tot}}{\partial z} \, .\\
\end{array}
\right.
\end{equation}
Here $\Phi_{\rm tot} = \Phi_{\rm disk} + \Phi_{\rm ext}$. 
Fig.~\ref{fig_FSLC.O_jeans} shows the radial profiles of
$\bar v_{\varphi}$, $\sigma_{\varphi}$, and $\sigma_z$
from the FSLC.O model and from the Jeans equations (\ref{eq_jeans})
(see also RS06).
It
can be seen that the model follows the Jeans equations very well. This is an
unexpected finding, especially considering such unusual moment profiles.

\begin{figure*}
\begin{center}
\includegraphics[width=16cm]{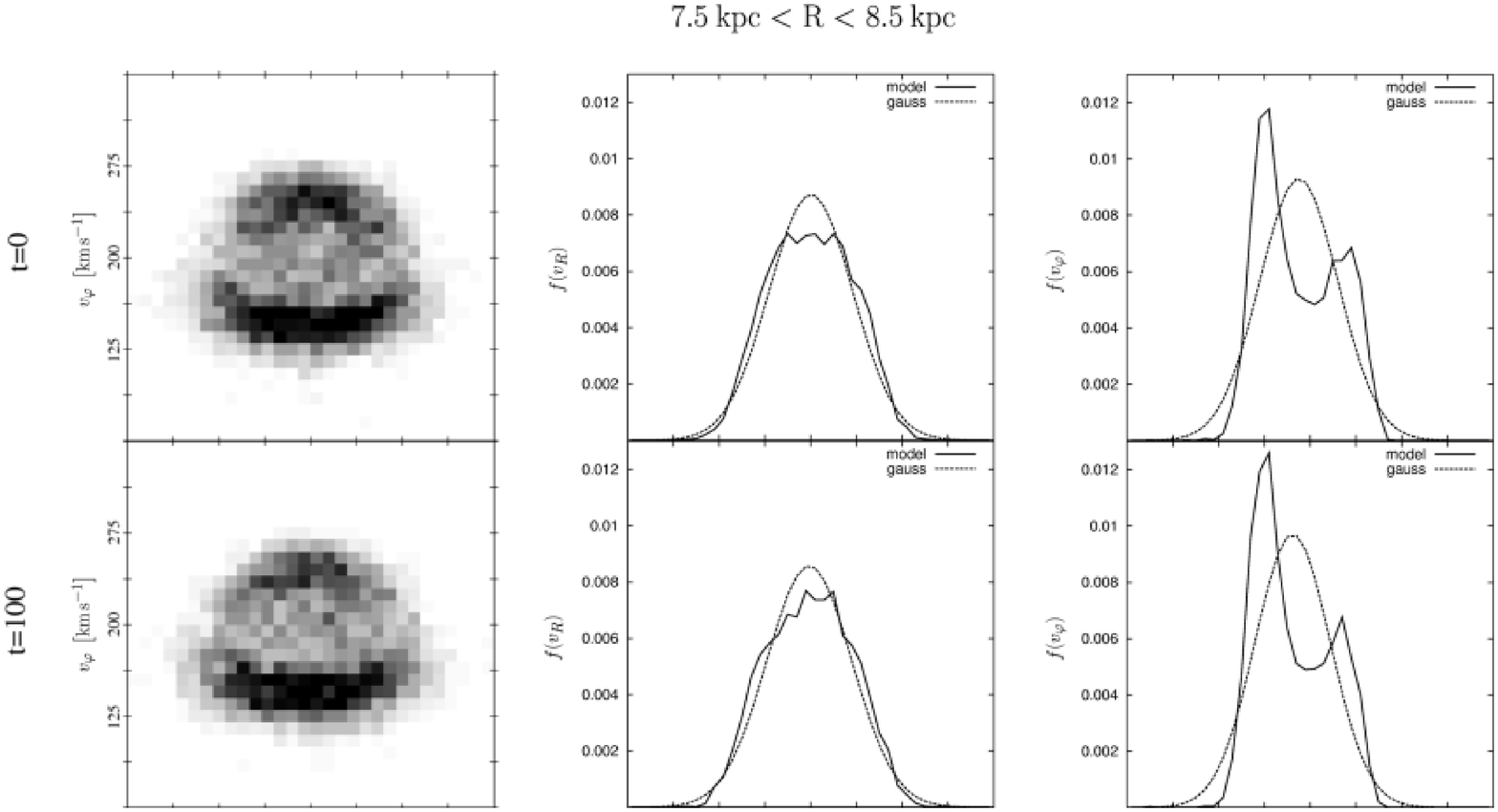}
\includegraphics[width=16cm]{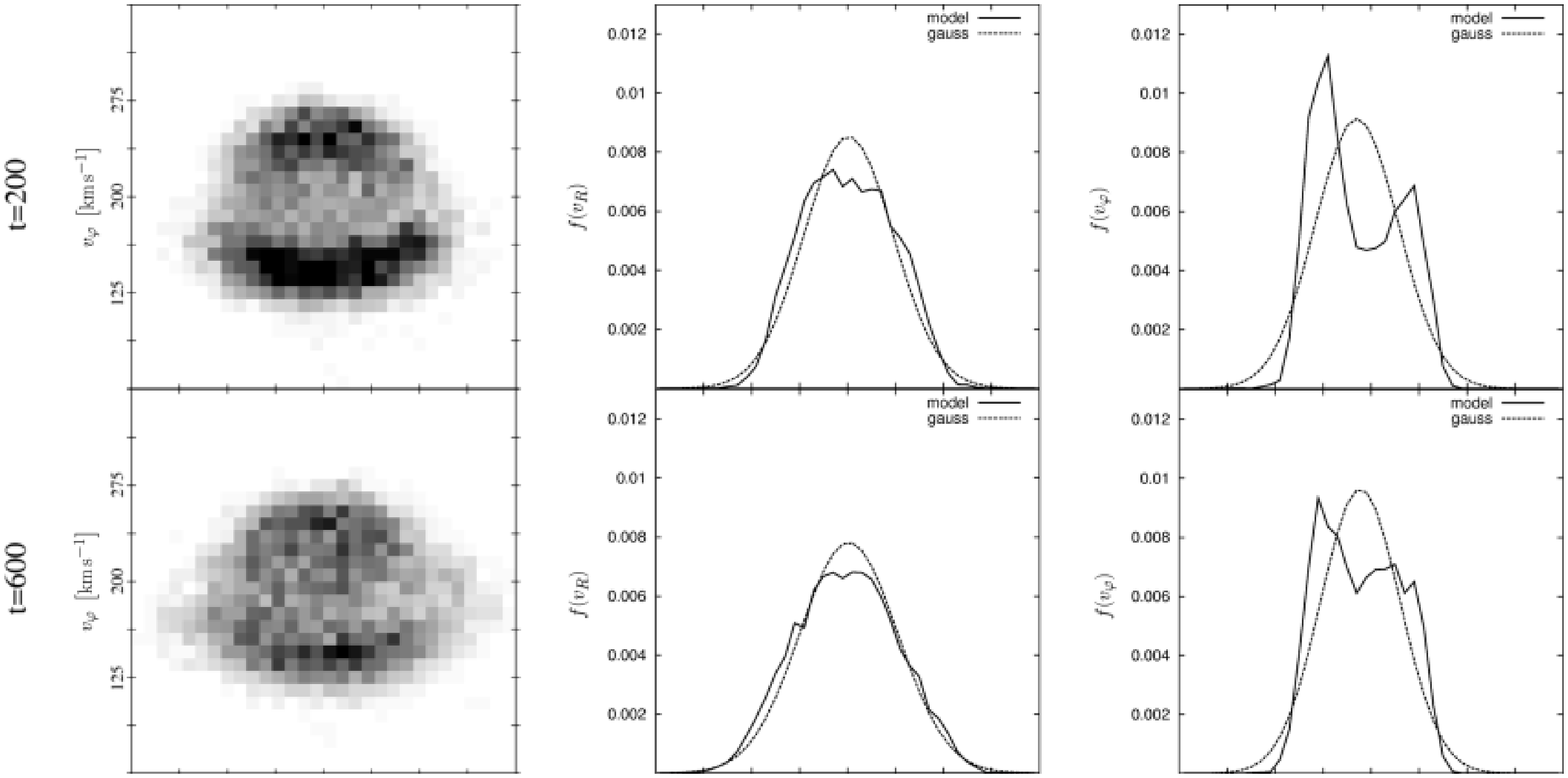}
\includegraphics[width=16cm]{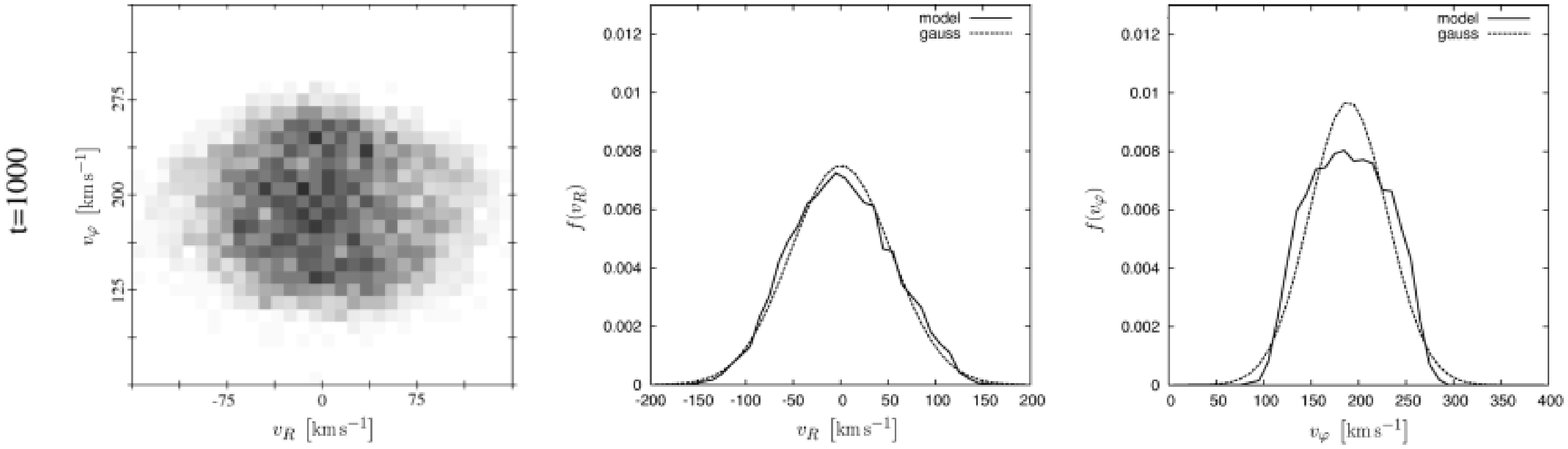}
\end{center}
\caption{
The velocity distributions for the FSLC.O model at various moments
of time (0, 100, 200, 600, and 1000~Myr). The region of the disc within the
range $7.5 \, {\rm kpc} < R < 8.5 \, {\rm kpc}$
is considered. Left column: two-dimensional
velocity distribution (the abscissa is the radial velocity $v_R$ and the
ordinate is the azimuthal velocity $v_{\varphi}$); 
the grey intensities correspond to
the numbers of particles that have velocities in the corresponding pixels.
Middle column: one-dimensional distribution of the velocity $v_R$. Right
column: one-dimensional distribution of the velocity $v_{\varphi}$. 
In the middle and
right columns, the solid lines show the model distributions, and the dashed
lines correspond to the Gaussians with parameters (mean and dispersion)
taken from the models.}
\label{fig_FSLC.O_velhist}
\end{figure*}

Another important feature of the FSLC.O model is that the velocity
distribution in this model is far from the Schwarzschild distribution. The
velocity distributions in the solar neighbourhood (near $R = 8 \; \rm kpc$)
are shown 
in Fig.~\ref{fig_FSLC.O_velhist}. Initially, both radial and azimuthal
velocity distributions are 
far from Gaussian, although such unusual distributions are more or less in
equilibrium. At least, they are conserved during the initial stages of
evolution. However, these distributions are unstable, and they change
substantially after about 500~Myr. After about 1~Gyr, the distributions are
smoothed and tend to Gaussians.

Above we have discussed the self-consistent
evolution of the FSLC.O model. It is also interesting, however, to examine a
non-self-consistent evolution of this model; that is, to calculate the
evolution of $N$ test particles in the total potential of the FSLC model. As
expected, the model practically does not change during such ``evolution'' (at
least on a time-scale of 10 Gyr). In particular, the density profile and
non-Schwarzschild velocity distribution do not change. Thus, this shows
again that this non-Schwarzschid velocity distribution is in equilibrium.

All models constructed using the Orbit.NB approach have the following
properties. They are close to equilibrium. Moreover, the velocity
distributions in the models are non-Schwarzschild ones and may have various
forms. However, the velocity distributions tend to the Schwarzschild
distribution during the dynamical evolution on a time-scale of 1~Gyr. Thus,
although these models are close to
equilibrium, they are probably non-physical because of their
non-Schwarzschild velocity distributions. The arguments are as follows.

\begin{figure*}
\begin{center}
\includegraphics[width=16cm]{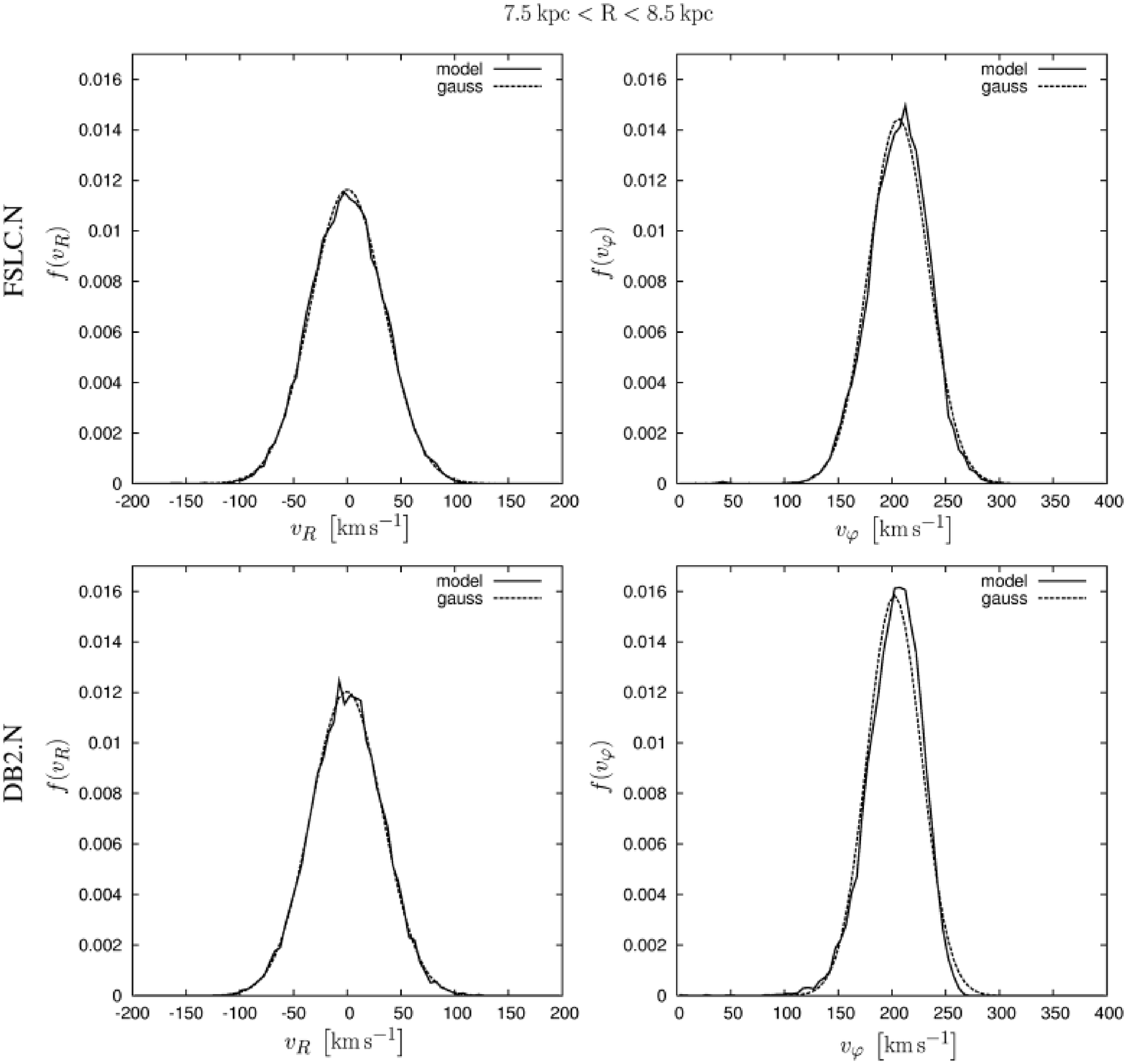}
\end{center}
\caption{
The distributions of radial and azimuthal velocities in the solar
neighbourhood in the models constructed using the Nbody.NB approach. We
consider the disc region of $7.5 \; {\rm kpc} < R < 8.5 \; {\rm kpc}$. 
Left panels: the
distributions of radial velocities. Right panels: the distributions of
azimuthal velocities. The upper pictures were constructed for the FSLC.N
model, and the lower ones for the DB2.N model. Description of both models is
given in Section~\ref{s_fsln.n_db2.n}. The solid lines show the model
distributions, and the 
dashed lines correspond to the Gaussians with parameters (mean and
dispersion) taken from the models.}
\label{fig_06}
\end{figure*}

\begin{itemize}
\item
The velocity distribution of the stars in the solar neighbourhood is similar
to the Schwarzschild distribution (see, for example, \citealt{BM}).
\item
The constructed non-Schwarzschild velocity distributions are
almost in equilibrium, but are unstable. The velocity distributions always
tend to the Schwarzschild one during the evolution.
\item 
In the models
constructed using the Nbody.NB approach, the final velocity distributions
are close to the Schwarzschild distribution (see Fig.~\ref{fig_06}). 
We note that the
Nbody.NB and Orbit.NB approaches differ only in the method of evolution
simulations in the iterations (see Section~\ref{s_evolv_method}).
\end{itemize}

Generally speaking, we
can assume that such unusual non-Schwarzschild velocity distributions will
survive only in the ``hothouse'' conditions of the Orbit.NB approach because
the evolution simulation in this approach is carried out
non-self-consistently (see Section~\ref{s_evolv_method}). 
When the conditions are close to
realistic (as in the Nbody.NB approach), such distributions are smoothed and
gradually converge to the Schwarzschild distribution.

\subsection{Uniqueness hypothesis}

In RS06, the authors formulated a hypothesis of uniqueness: not
more than one equilibrium model (one equilibrium distribution function) can
exist for a fixed density $\rho_{\rm disk}(R,z)$ and 
potential $\Phi_{\rm ext}(R,z)$ and a fixed
kinetic energy fraction of residual motions (e.g. a fixed angular momentum
$L_z$).

We can now say that in such a form the hypothesis is false. We can
construct using the Orbit.NB approach as many equilibrium models as we want
for the same $\rho_{\rm disk}(R,z)$, $\Phi_{\rm ext}(R,z)$ and fixed $L_z$. 
However, although these
models are close to equilibrium, they probably bear no relation to actual
stellar systems.

At the same time, the models constructed by the Nbody.NB
approach are the same (within the noise level) for arbitrary initial states
at the same $\rho_{\rm disk}(R,z)$, $\Phi_{\rm ext}(R,z)$ and fixed $L_z$. 
Moreover, the velocity
distributions in constructed models are always close to the Schwarzschild
distribution. Based on this fact and the probable non-physical character of
the Orbit.NB models, one can formulate a hypothesis that the ``physical''
discs in the equilibrium state are unique. This hypothesis could be
formulated as follows. When the functions $\rho_{\rm disk}(R,z)$ and
$\Phi_{\rm ext}(R,z)$ are
fixed and a fraction of the kinetic energy in the residual motions is also
fixed (e.g. the value of the angular momentum $L_z$ is fixed), not more than
one physical model in the equilibrium state may exist (the physical model
is the model that can exist in reality). We think that the main feature of a
real stellar disc is an almost Schwarzschild velocity distribution.

Whether
or not our hypothesis is true is very important. If we know the mass
distribution in the Galaxy and our hypothesis is
valid, we can use our Nbody.NB method to reconstruct the velocity
distribution in the Galactic disc. In other words, we can construct a phase
model of the Galactic disc, and this model will correspond exactly to the
real Galactic disc.
 
\section{Phase models of the galactic stellar disk}
\label{s_phase_model}

\subsection{Choice of model among the family of models}

For the construction of
phase models of the Milky Way Galactic disc, we use the Nbody.NB approach
(see Section~\ref{s_itdisk}). Using this method, one can construct a family of models
for the fixed functions $\rho_{\rm disk}(R,z)$ and $\Phi_{\rm ext}(R,z)$. 
This is a one-parameter
family, and the parameter is the fraction of kinetic energy of residual
motions (in other words, the disc ``heat'' degree). In our approach, this
parameter is the value of $L_z$. We emphasize that, using the Nbody.NB
approach, when we fix the value of $L_z$ we obtain the same model (within the
noise level) regardless of the initial state.

\begin{figure*}
\begin{center}
\includegraphics[width=16cm]{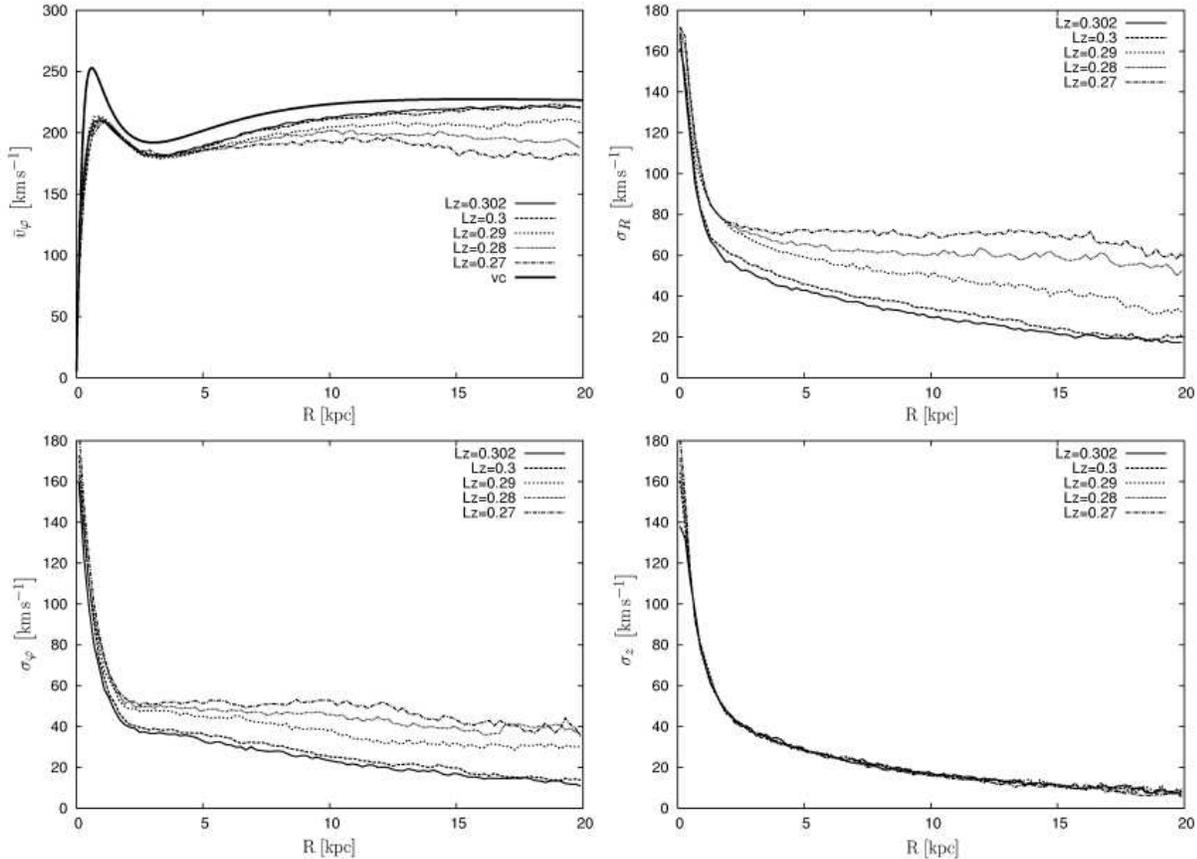}
\end{center}
\caption{
The family of phase models constructed using the Nbody.NB approach
for the FSLC density model. For each model, we show the dependences of 
$\bar v_{\varphi}$, $\sigma_R$, $\sigma_{\varphi}$, and $\sigma_z$
on $R$.
The moments were calculated in cylindrical layers
infinite along the $z$-axis. In the upper left panel, the circular velocity
curve is shown by the thick solid line. Models constructed for 
$L_z = 0.302, \, 0.3, \, 0.29, \, 0.28, \, 0.27$ are represented. 
The model with $L_z = 0.302$ is
the FSLC.N model corresponding to the Galactic stellar disc.}
\label{fig_fslc_family}
\end{figure*}

A family of the $N$-body models
for the FSLC density model is shown in Fig.~\ref{fig_fslc_family} as 
an example. The parameters
of the models are given below. One can ask how we choose the best-fitting
model among the family. We have comparatively reliable kinematic Galactic
parameters in the solar neighbourhood (see, for example, \citealt{BM, DBb}),
so it is reasonable to use them to choose the
model. One needs to choose any one parameter that on the one hand is well
known in the solar neighbourhood, and on the other depends strongly on the
disc ``heat''. In other words, this value has to be strongly dependent on the
value of $L_z$. For example, the velocity ellipsoid parameters 
$\bar v_{\varphi}$, $\sigma_R$, $\sigma_{\varphi}$, and $\sigma_z$
are well known. The value $\sigma_z$ is not suitable, 
because it does not depend
on $L_z$ (see Fig.~\ref{fig_fslc_family} and RS06). 
The value $\bar v_{\varphi}$ is also unsuitable, because it
depends weakly on $L_z$ for cold models (see Fig.~\ref{fig_fslc_family}).
The choice between the
two values $\sigma_R$ and $\sigma_{\varphi}$ is somewhat arbitrary. 
We prefer the value $\sigma_R$ for the model choice.

There are various estimates of the value of $\sigma_R$ in the
literature. We have chosen the value 
$\sigma_R = 35 \; {\rm km} \, {\rm s}^{-1}$ that was estimated using
an extrapolation to the zero heliocentric distance (see \citealt{OMSO}).
In addition, we adopted the solar distance from the Galactic centre as 
$R_0 = 8 \; {\rm kpc}$
(see, for example, \citealt{N, A}).

As a result, we have chosen the model in which the radial velocity
dispersion in the solar neighbourhood is about 
$35 \; {\rm km} \, {\rm s}^{-1}$. As the solar
neighbourhood, we have chosen the region 
$7.9 \; {\rm kpc} < R < 8.1 \; {\rm kpc}$ and $|z| < 0.1 \; {\rm kpc}$.

\subsection{Models FSLC.N and DB2.N}
\label{s_fsln.n_db2.n}

We have considered two families of stellar
disc models constructed for two Galactic density models (FSLC and DB2). The
families of
phase models were constructed using the Nbody.NB approach. From every
family, we have chosen the stellar disc model that corresponds to the
Galactic disc in the solar neighbourhood (in terms of the radial velocity
dispersion). Hereafter we refer to these models as FSLC.N and DB2.N.

The
family of models for the FSLC density model is shown in
Fig.~\ref{fig_fslc_family}. It was
constructed in the following way. We took as $\rho_{\rm disk}$ 
the disc density
distribution from the FSLC model, and as $\Phi_{\rm ext}$ the potential 
arising from all
FSLC model components except the disc. The disc density was taken as equal
to zero at 
$R > R_{\rm max} = 30 \; {\rm kpc}$ or $|z| > z_{\rm max} = 10 \; {\rm kpc}$.
We considered an
initial cold model in which the particle orbits are circular. Four iteration
sets with increasing accuracies were calculated. The parameters of the sets
are shown in Table~\ref{tab_nn}. The number of neighbours in the 
velocity distribution
transfer is $n_{\rm nb} = 10$. The time of one iteration is
$t_i = 50 \; {\rm Myr}$. The FSLC.N
model was constructed for the angular momentum $L_z=0.302$ (this value is
given in our system of units as described in 
Section~\ref{s_strange_model_s1}).

\begin{table}
\caption{
The parameters of four iteration sets for the family of models
constructed using the Nbody.NB approach for the FSLC and DB2 density models.
Here, $n_{\rm it}$ is the number of iterations, $N$ is the number of bodies, 
$dt$ is the
integration step, $\epsilon_f$ is the softening length for the FSLC model, 
and $\epsilon_d$ is the
softening length for the DB2 model. The softening length was chosen using
the recommendations of \citet{RS05}.}
\label{tab_nn}
\begin{center}
\begin{tabular}{c|cccc}
\hline
$n_{\rm it}$ & N &  $dt, \; {\rm Myr}$ & $\epsilon_f, \; {\rm kpc}$ & 
$\epsilon_d, \; {\rm kpc}$\\
\hline
500 & 20,000  & 1    & 0.04 & 0.04 \\
200 & 100,000 & 0.5  & 0.02 & 0.02 \\
100 & 500,000 & 0.5  & 0.02 & 0.01 \\
5   & 500,000 & 0.05 & 0.02 & 0.01 \\
\hline
\end{tabular}
\end{center}
\end{table}

The family of
models for the DB2 density model, in particular the DB2.N model, was
constructed in the following way. The density of the thin stellar disc in
the DB2 model was taken as $\rho_{\rm disk}$,
and the potential arising from all
components except the thin stellar disc was taken as $\Phi_{\rm ext}$.
The disc density
at $R > R_{\rm max} = 30 \; {\rm kpc}$
or $|z| > z_{\rm max} = 10 \; {\rm kpc}$ was taken as equal to zero. This
is the same condition as in the FSLC~model. Here we again consider the cold
initial model with circular orbits. Four sets of iterations were again
calculated. The parameters of these sets are shown in Table 3. We have
adopted the parameters $n_{\rm nb} = 10$, $t_i = 50 \; {\rm Myr}$.
The DB2.N model was
constructed for the angular momentum $L_z=0.1595$ 
(this value is given in our
system of units as described in Section~\ref{s_strange_model_s1}).

\begin{figure*}
\begin{center}
\includegraphics[width=16cm]{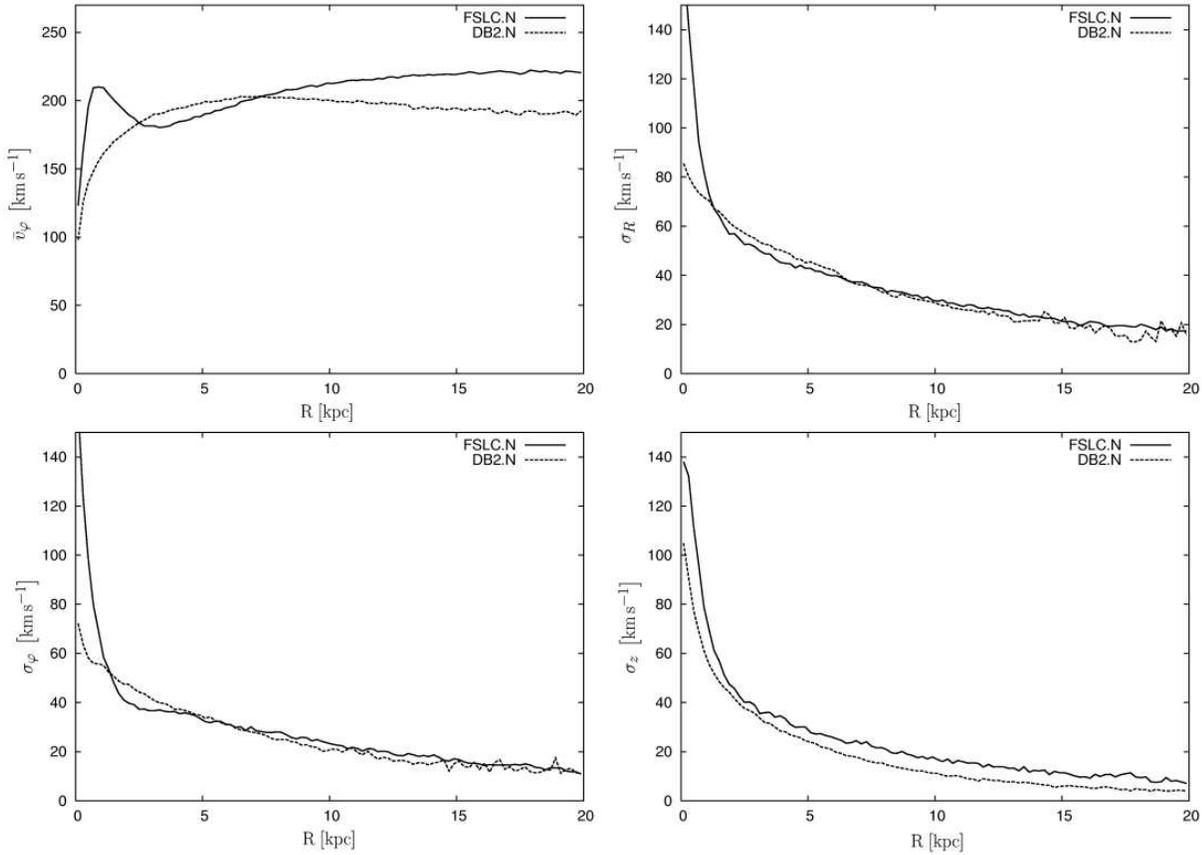}
\end{center}
\caption{
The dependences of four velocity distribution moments 
$\bar v_{\varphi}$, $\sigma_R$, $\sigma_{\varphi}$, and $\sigma_z$
on the cylindrical radius $R$ for the FSLC.N and DB2.N models. The
moments were calculated in cylindrical layers infinite along the $z$-axis.
}
\label{fig_N_moments}
\end{figure*}

The radial profiles of the
velocity distribution moments for the FSLC.N and DB2.N models are shown in
Fig.~\ref{fig_N_moments}. Let us consider the profiles 
of $\sigma_R$ and $\sigma_{\varphi}$. In the central parts of
the models, these profiles are very different. This is caused by the more
massive and concentrated bulge in the FSLC.N model with respect to the DB2.N
model (see Fig.~\ref{fig_densmod} in Section~\ref{s_densmod}).
However, the profiles of $\sigma_R$ and $\sigma_{\varphi}$ for the
two models are similar from about $2-3$~kpc. This is in spite of the
difference between the initial density models. In particular, one can
observe a different relative disc contribution in the total mass and
rotation curve (see Fig.~\ref{fig_densmod}). Fig.~\ref{fig_N_moments} 
also shows that the profiles of $\sigma_z$ are
different. However, we note that the value of $\sigma_z$ at any point 
is defined
only by the density distribution. This fact is a consequence of the last
Jeans equation (\ref{eq_jeans}). Our phase models satisfy the Jeans
equations very well, 
however, and therefore the differences in the $\sigma_z$ profiles
are explained by
the differences in the density models.

\begin{figure*}
\begin{center}
\includegraphics[width=16cm]{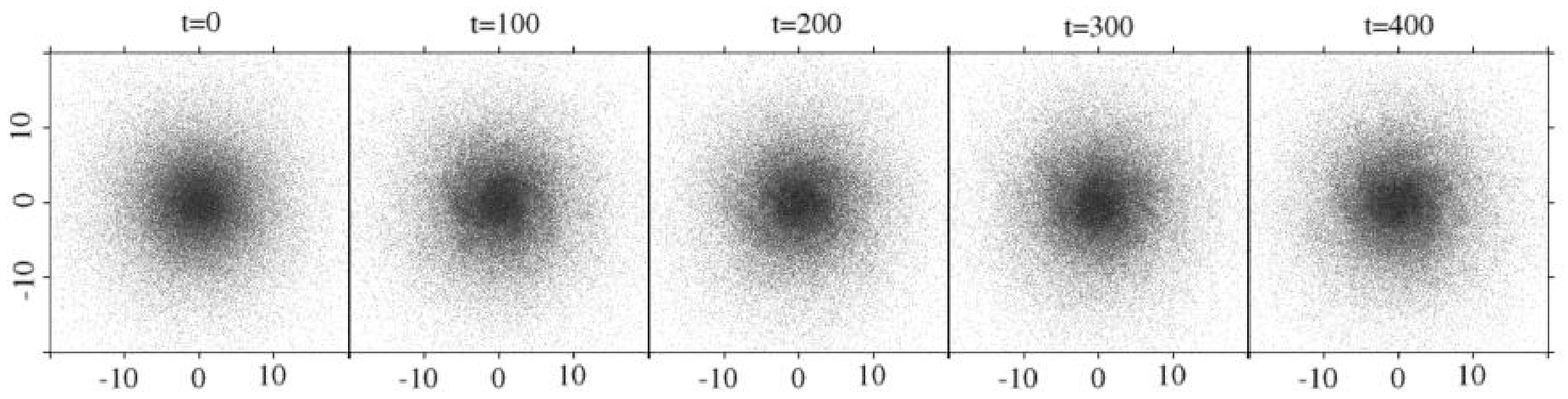}
\includegraphics[width=16cm]{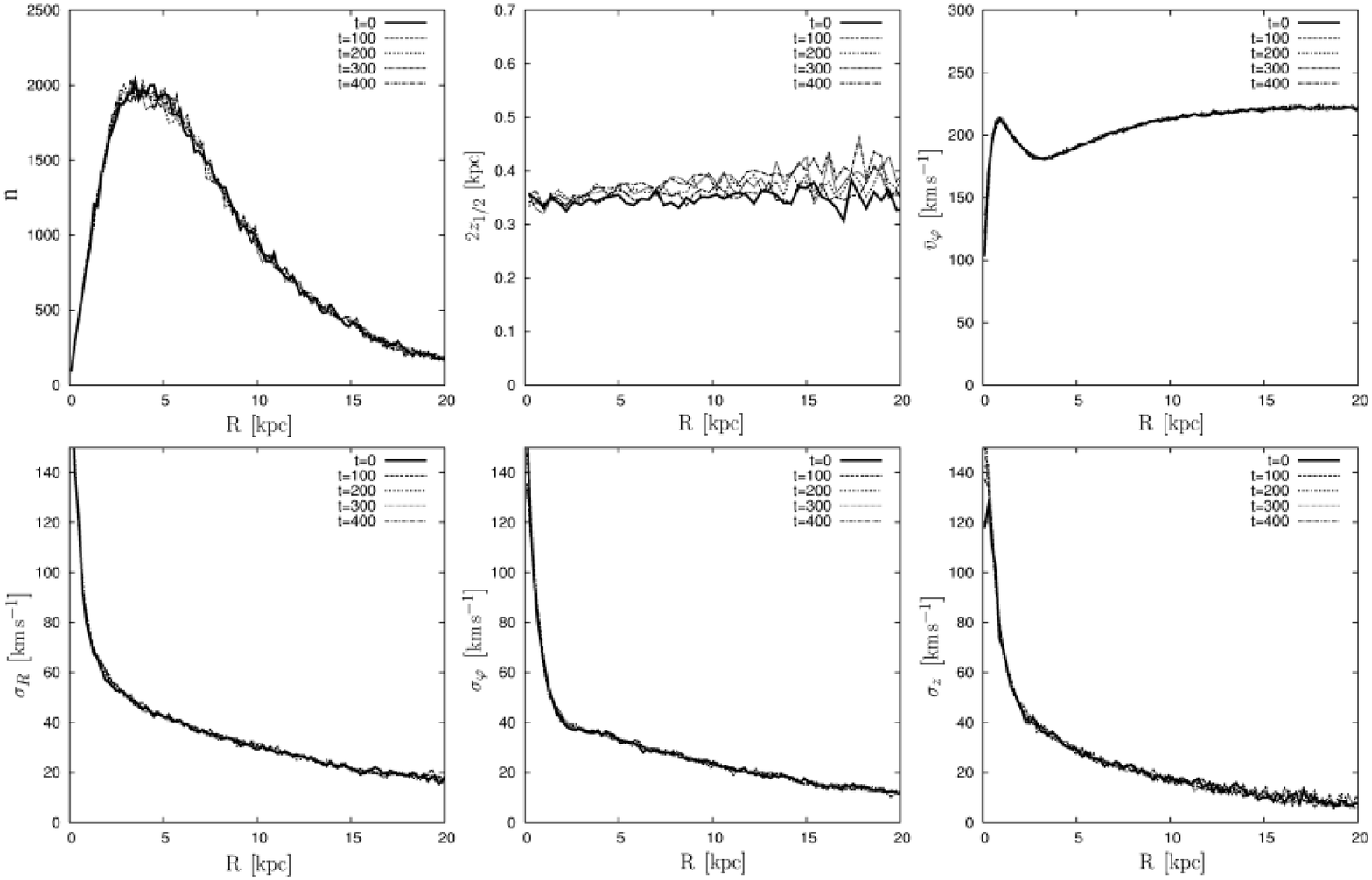}
\end{center}
\caption{
Initial evolution stages of the FSLC.N model. The same values are
shown as in Fig.~\ref{fig_FSLC.O_stability}. 
The evolution simulation used the following parameters:
number of bodies $N=100\,000$, integration step $dt=0.04 \; \rm Myr$,
softening length $\epsilon=0.025 \; \rm kpc$. The two last parameters were
chosen according to the 
recommendations of \citet{RS05}.}
\label{fig_fslc_n_stability}
\end{figure*}

\begin{figure*}
\begin{center}
\includegraphics[width=16cm]{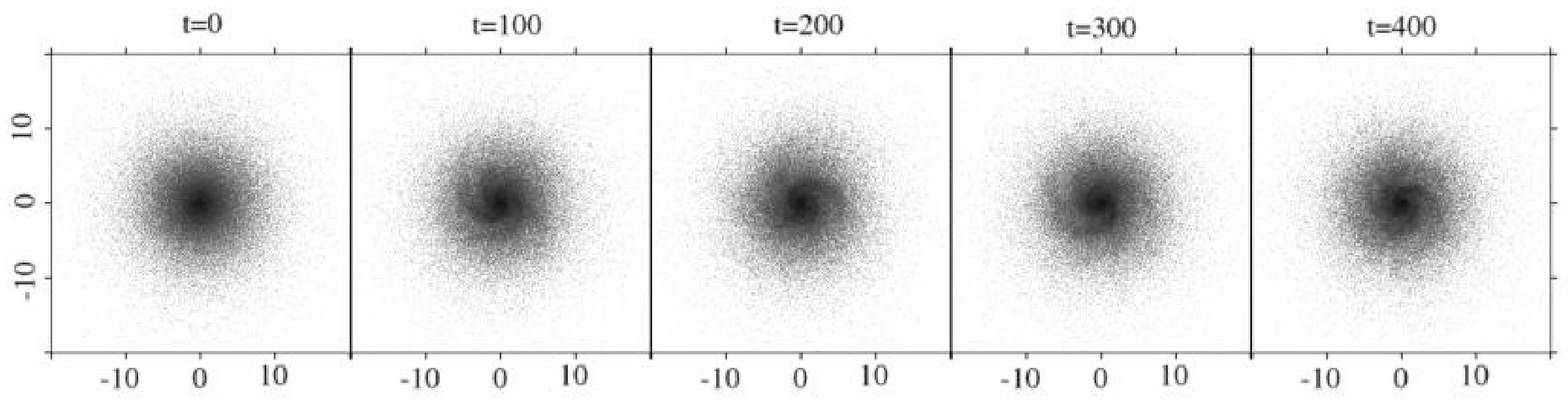}
\includegraphics[width=16cm]{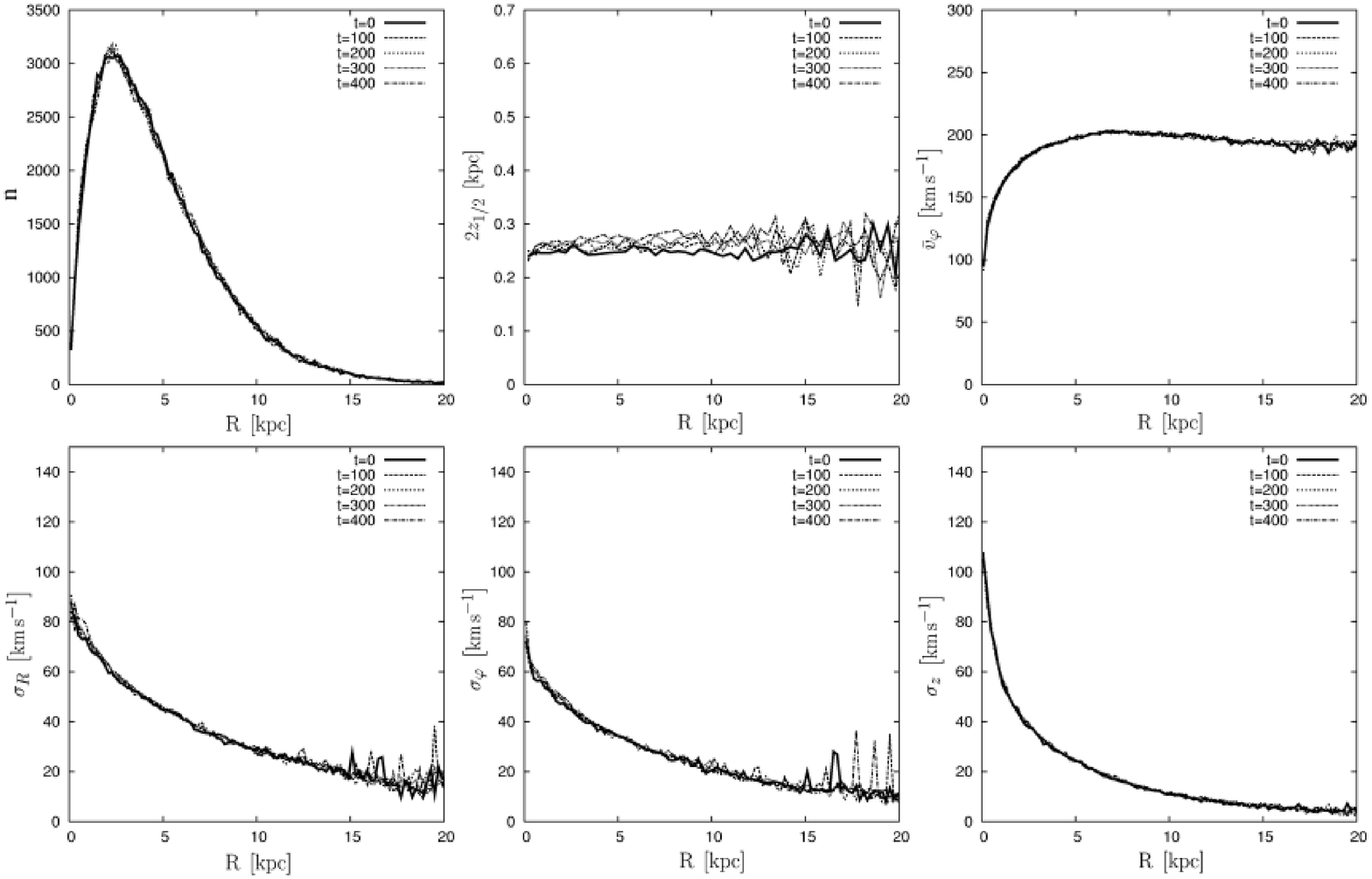}
\end{center}
\caption{
Initial evolution stages of the DB2.N model. The same values are
shown as in Fig.~\ref{fig_FSLC.O_stability}. The evolution simulation used
the following parameters: 
number of bodies $N=100\,000$, integration step $dt=0.04 \; \rm Myr$,
softening length $\epsilon=0.015 \; \rm kpc$. 
The two last parameters were chosen according to the
recommendations of \citet{RS05}.}
\label{fig_db2_n_stability}
\end{figure*}

The initial stages of the evolution for the FSLC.N and DB2.N models are
shown in Figs~\ref{fig_fslc_n_stability}~and~\ref{fig_db2_n_stability}. 
Both models conserve the structural and dynamical
parameters in the early stages of the evolution very well. Therefore both
models are close to equilibrium.

\section{Conclusions}
\label{s_conclusions}

In this paper, we have discussed the problem of constructing
a phase model of the Galactic stellar disc. We used the iterative method
proposed in RS06. The realization of the iterative method described in RS06
(Nbody.SCH approach) has some disadvantages. The Nbody.SCH method has a
problem with the construction of a relatively cold stellar disc.
Furthermore, the Nbody.SCH method cannot be directly applied to stellar
systems with arbitrary geometry. The main goal of this study was to develop
a new realization of the iterative method without the disadvantages of the
Nbody.SCH approach. For this purpose, we considered a number of
modifications of the iterative method (Nbody.SCH, Orbit.SCH, Nbody.NB, and
Orbit.NB). The Nbody.NB method satisfied our conditions. This method can be
directly applied for the construction of phase models of stellar systems
with arbitrary geometry. The method is simple in terms both of understanding
and of implementation. Using the Nbody.NB approach, we have constructed
phase models of the Galactic disc for two realistic density models
(suggested by \citealt{FSLC, DBa}).

For a given mass distribution model of the Galaxy, we can construct a
one-parameter family of equilibrium models of the Galactic disc. From this
family we can choose a unique model using local kinematic parameters that
are known from the Hipparcos data. There are, however, two important
questions. Is there an equilibrium disc model besides the models from this
one-parameter family? Does the real Galactic disc belong to this
one-parameter family? (Of course this family should be constructed for a
real mass model of the Galaxy.) The answer to the first question is
affirmative. Using the Orbit.NB method, we can construct many models besides
the models from this one-parameter family. We have, however, shown that
these models are probably non-physical. Based on this fact, we suppose that
all models except models from this one-parameter
family are non-physical. Consequently, we think that the answer to the
second question is also affirmative. However, we cannot strictly prove it as
yet. We think that the key to the proof is Schwarzschild velocity
distributions. The models from our one-parameter family have almost such
velocity distributions, and we think that the velocity distributions in real
galactic discs are close to the Schwarzschild distribution.

We now discuss
possible applications of our Nbody.NB method.

This method can be used to
construct phase models of stellar systems with arbitrary geometry. For
example, the method can be used to construct phase models of triaxial
elliptical galaxies.

Here we have used the iterative method to construct a
phase model of the stellar disc of a spiral galaxy. In the future, we are
going to construct a self-consistent model of a spiral galaxy (including
live halo and bulge) using a modified Nbody.NB approach.

Another, rather more important, direction of future investigations is a
comparison of our iterative models with observations of real galaxies. This
comparison will make it possible to derive from observations the constraints
on unobservable parameters of galaxies (for example dark matter mass and its
profile). Of course, such a comparison should first be carried out for the
Milky Way. One of the possible schemes is as follows. If we know the mass
distribution in the Galaxy then using our method we can construct a phase
model of the Galactic disc. The mass distribution in the Galaxy contains two
parts --- visible and dark matter. From the phase model of the Galactic disc,
we can derive the profiles of stellar kinematic parameters (profiles of
velocity dispersions and mean azimuthal velocity). Let us assume that we
know from observations the mass distribution of visible matter and the
profiles of stellar kinematic parameters. Then, using the iterative method,
we can put some constraints on the dark
matter mass distribution. The distribution of the dark matter should be such
that the iterative model has the observable kinematics. At the moment,
observational profiles of stellar kinematic parameters have fairly large
uncertainties; however, it is expected that the GAIA mission will provide a
much higher accuracy for these data. We are planning to study which
constraints on the Milky Way parameters we can obtain from GAIA using our
iterative models.

The iterative method can also be used to derive the
velocity dispersion profiles from observations. It is impossible to obtain
the velocity dispersion profiles for external spiral galaxies directly from
observations, which can yield only the line-of-sight velocity, $v_{\rm los}$,
and the line-of-sight velocity dispersion, $\sigma_{\rm los}$.
However, we can use the Jeans
equations to derive the velocity dispersion profiles from the observed
quantities  $v_{\rm los}$ and $\sigma_{\rm los}$. 
In practice, however, one has to include some {\it a priori} 
assumptions about the form of the velocity
dispersion profiles. In the literature, authors have made slightly different
assumptions, but all of them are based on the hypothesis that the velocity
dispersion in the radial direction is proportional to the velocity
dispersion in the vertical direction (see section 3 in RS06).

Using our
iterative method, we can derive the velocity dispersion profiles from
line-of-sight parameters without any additional assumptions. The general
idea is as follows. From observations we have (more or less precisely) the
distribution of visible mass and some constraints on the distribution of
dark matter. If we fix the mass distribution in a galaxy (for visible and
dark matter) then, using the iterative method, we can construct a
one-parameter family of phase models of the galactic disc. The parameter of
this family is the degree of disc heating (for example the fraction of the
kinetic energy of the disc contained in residual motions). Using observable
profiles of $v_{\rm los}$ and
$\sigma_{\rm los}$, we can choose a unique model from this family.
As a result, we will have a phase model of the galactic disc. From this
model we can calculate the velocity dispersion profiles.

\section*{Acknowledgements}

We thank Drs Nataliya Sotnikova and Mariya Kudryashova for many helpful
comments. 

We thank for financial support the government of Saint-Petersburg
(grant PD07-1.9-51), Russian Foundation for Basic Research (grant
06-02-16459), and Leading Scientific School (grant NSh-8542.2006.02 and
grant NSh-4929.2006.02).

\label{lastpage}
\end{document}